\documentclass[manuscript,screen, authorversion,nonacm]{acmart}

\AtBeginDocument{%
  }

\setcopyright{acmlicensed}
\copyrightyear{2025}
\acmYear{2025}
\acmDOI{XXXXXXX.XXXXXXX}

\acmConference[Conference acronym 'XX]{Make sure to enter the correct conference title from your rights confirmation emai}{June 03--05,  2018}{Woodstock, NY}

\setcopyright{none}
\renewcommand\footnotetextcopyrightpermission[1]{}
\usepackage{dcolumn}
\newcolumntype{d}{D{.}{.}{2}}

\acmISBN{978-1-4503-XXXX-X/18/06}

\begin{document}



\title["How to Explore Biases in Speech Emotion AI with Users?"]{"How to Explore Biases in Speech Emotion AI with Users?" \protect\\ A Speech-Emotion-Acting Study Exploring Age and Language Biases}





\author{Josephine Beatrice Skovbo Borre}
\email{jborre16@student.aau.dk}
\orcid{}
\affiliation{%
  \institution{Aalborg University}
  \department{Department of Computer Science}
  \city{Aalborg}
  \country{Denmark}
}

\author{Malene Gorm Wold}
\email{mwold23@student.aau.dk}
\orcid{}
\affiliation{%
  \institution{Aalborg University}
  \department{Department of Computer Science}
  \city{Aalborg}
  \country{Denmark}
}

\author{Sara Kjær Rasmussen}
\email{skra23@student.aau.dk}
\orcid{}
\affiliation{%
  \institution{Aalborg University}
  \department{Department of Computer Science}
  \city{Aalborg}
  \country{Denmark}
}

\author{Ilhan Aslan}
\email{ilas@cs.aau.dk}
\orcid{0000-0002-4803-1290}
\affiliation{%
  \institution{Aalborg University}
  \department{Department of Computer Science}
  \city{Aalborg}
  \country{Denmark}
}

\renewcommand{\shortauthors}{}


\begin{abstract}
This study explores how age and language shape the deliberate vocal expression of emotion, addressing underexplored user groups, Teenagers (\textit{N}~=~12)  and Adults 55+ (\textit{N}~=~12), within speech emotion recognition (SER). While most SER systems are trained on spontaneous, monolingual English data, our research evaluates how such models interpret intentionally performed emotional speech across age groups and languages (Danish and English). To support this, we developed a novel experimental paradigm combining a custom user interface with a backend for real-time SER prediction and data logging. Participants were prompted to ``hit'' visual targets in valence-arousal space by deliberately expressing four emotion targets. While limitations include some reliance on self-managed voice recordings and inconsistent task execution, the results suggest contrary to expectations, no significant differences between language or age groups, and a degree of cross-linguistic and age robustness in model interpretation. Though some limitations in high-arousal emotion recognition were evident.
Our qualitative findings highlight the need to move beyond system-centered accuracy metrics and embrace more inclusive, human-centered SER models. By framing emotional expression as a goal-directed act and logging the real-time gap between human intent and machine interpretation, we expose the risks of affective misalignment. 
\end{abstract}






\maketitle

\section{Introduction}

As speech emotion recognition (SER) systems become
increasingly embedded in domains such as mental health
support, education and human-machine interaction, the demand for inclusive and ethically grounded affective computing grows more urgent \cite{29_mayong2024, 6_aslan2025speejis, 3_AnPengcheng2024, 21_hansen2022generalizable, 2_Adeleye2024}. While technical advancements in deep learning architectures for computer audition, such as Wav2Vec 2.0, have driven model performance, much of the field continues to rely on monolingual, English-centric datasets and spontaneous emotional expression, often overlooking the diversity of human users \cite{38_Wagner23, 13_dair2022, 28_Luo2024, 10_catania2023speech,40_wang2025, 41_Wani}. This paper addresses several of these gaps by exploring how
age and language shape users’ ability to deliberately express
emotion through speech. Specifically, we investigate how
Teenagers vs Adults 55+ perform when tasked with
intentionally expressing four different emotions, and how a pre-trained SER model from Wagner et al. \cite{38_Wagner23} interprets those vocal expressions, particularly from native Danish speakers, across two distinct age groups.
A key innovation of this study lies in its experimental
design. We developed a custom Processing-based user
interface (UI) inspired by emotion targeting mechanics,
where participants were invited to “hit” a visualised target
coordinate in valence-arousal (V A) space using only their
voice. The method is inspired by research and design tools such as Fitts's Law \cite{mackenzie1992fitts} and elicitation studies with users \cite{Wobbrock09}, which have also been used for contextual elicitation \cite{contextual_elicitation_18}, asking users to perform fitting actions for a predefined target or effect. 

This front-end was integrated with a Python-based backend that handled real-time emotion prediction, transcription, and logging, allowing us to continuously monitor how emotional intent was expressed and interpreted during the experiment.
To this end, we build on the architecture of previous work \cite{5_aslan2025speechcommand}. By transforming emotional expression into a goal-directed task, we captured rich, real-time data on how emotion is intentionally performed, and misinterpreted, across demographic and linguistic boundaries. Our approach shifts the analytical lens from system-level recognition accuracy to human-level expressive accuracy. It also questions whether current SER models, trained on predominantly English-language corpora, can generalise across typologically different languages like Danish, and whether age-related patterns influence emotional clarity in model interpretation.

Ultimately, this research contributes to the broader SER field by advocating for a human-centered, inclusive perspective. Rather than asking users to adapt their expressiveness to system expectations, we invert the relationship: how can SER systems better attune to the variability of human emotion? By focusing on deliberate expression and age/language diversity, our findings offer not
only technical insight but also ethical grounding, moving toward emotionally intelligent technologies that reflect the full spectrum of human affect. Without such consideration, we risk building systems in which emotions, and the people behind them, are, quite literally, lost in translation between human and machine.

\section{Background}

\subsection{Speech emotion Recognition (SER) and emotion
expression in speech}
People's ability to express emotional states through speech is a foundational component of human communication, particularly in voice-only contexts where visual cues like facial expressions and body language are absent. Emotional states can be identified not only from speech content (what is said) but also from linguistic and paralinguistic cues (how it is said) \cite{25_kraus2017voice, 6_aslan2025speejis}. Building on these foundations, classical automatic SER have emerged within human-computer interaction (HCI) \cite{41_Wani,24_Khalil}. SER systems are designed to process voice signals and evaluate emotional states or dimensional value in human-to-machine interactions \cite{38_Wagner23}. Wagner et al. proposed that research in SER is dominated by two conceptual paradigms (i) discrete emotions (e.g. happy and
sad), and (ii) underlying dimensions (arousal, valence, and dominance). Early research in SER often relied on category (i), namely the labelling of emotions \cite{8_Bestelmeyer17}. For instance, Ekman's emotional model focuses on identifying discrete emotions such as happiness, sadness, anger and calmness \cite{3_AnPengcheng2024, 15_ekman1992there, 16_Ekman}, whereas the Circumplex Model of Affect, introduced by Russell and Barrett offers a way to position each emotion within a two-dimensional (2D) model \cite{37}. Emotions such as happy are not limited to one single point within the Circumplex Model of Affect but are distributed across a range within the valence-arousal space \cite{26_kutsuzawa2022classification}. Valence refers to the quality of an emotion, distinguishing between negative and positive, while arousal refers to the intensity of an emotion, ranging from low-arousal (calm) to high-arousal (excited) \cite{4_AN2025,8_Bestelmeyer17}.

\subsection{Speech models}
The integration of valence-arousal models within SER provides a more nuanced and multidimensional method for identifying emotional states in speech \cite{36_rizhinashvili2024enhanced}. James et al.~\cite{23_james5063641emotiongui} emphasised the widespread acceptance of Russell’s V A model as a foundational framework in numerous two-dimensional speech emotion models. To increase the precision of emotion categorisation,
SER models based on Wav2vec 2.0 are generally pre-trained on large-scale datasets before being fine-tuned on smaller, annotated datasets \cite{38_Wagner23}. Originally created by Baevski et al., Wav2vec 2.0 is a self-supervised learning model utilising a Transformer architecture, learning speech features from raw audio \cite{7_baevski2020wav2vec20}. Since its introduction, Wav2vec 2.0 has been fine-tuned on diverse datasets including MSP-Podcast, IEMOCAP, and EmoDB to classify emotions in both acted and naturalistic speech \cite{39_wang2022finetunedwav2vec20, 38_Wagner23}. Several
studies have extended its application beyond typical use-cases. Beyond standard datasets, Catania and Garzotto \cite{10_catania2023speech} proposed Emozionalmente, a novel model using Wav2vec 2.0 to address linguistic limitations in existing corpora. They developed a custom dataset with 6,902
samples recorded of 431 Italian speakers to improve emotion recognition in underrepresented languages \cite{10_catania2023speech}. In a separate context, Speejis \cite{6_aslan2025speejis} used the SER model from Wagner et al. \cite{38_Wagner23} to explore emotional cues in augmented voice messaging, integrating emojis into spoken interactions to increase expressivity. These studies demonstrate Wav2vec 2.0’s ability to adapt to a wide range of SER
implementations.

\subsection{Bias in SER models: Age and Language Inclusivity}
SER models built on existing pre-trained datasets are vulnerable to bias stemming from imbalances in speaker characteristics such as gender, age, and linguistic or cultural background \cite{36_rizhinashvili2024enhanced, 10_catania2023speech, 28_Luo2024}. Datasets are typically programmed in one language and developed to detect valence-arousal dimensions from the most general emotions: anger, disgust, fear, joy, sadness, surprise and neutrality \cite{10_catania2023speech}. Models trained on standardised or dominant language forms often struggle to perform accurately across varied linguistic profiles, thereby restricting their applicability in multilingual settings \cite{13_dair2022}. Such limitations can lead to biased outcomes, where errors in classifying specific
linguistic groups may misrepresent emotional expression \cite{33_Neumann}, undermining user trust and long-term use \cite{13_dair2022}. Age related bias is a particularly under examined dimension in
SER \cite{9_cahyawijaya2023}. A study by Park et al. \cite{34_Park21} revealed that older adults (aged 65+) are significantly underrepresented in large-scaled datasets (only five out of 92 datasets included
this age group). The SER model from Wagner et al. is trained on the MSP-Podcast corpus, although the dataset labels are not publicly available \cite{14_duret2024}. This potential lack of
representation raises concerns regarding individual fairness in model performance, as SER systems may be less accurate when interpreting emotional expressions from underrepresented demographics. Another source of bias in SER models relates to how emotional labels are applied and interpreted \cite{41_Wani}. Spontaneous speech is contextually richer, more ambiguous or with blended emotional states which can be harder to label than fixed sentences \cite{17_2024emovomedataset}. As a result,
models trained exclusively on fixed-sentences and acted datasets may also underperform when applied to
spontaneous settings \cite{17_2024emovomedataset}. This gap might further affect groups whose emotional expressiveness is subtler, including older adults or speakers for certain cultural backgrounds, and thereby compounding demographic and representational biases in SER systems \cite{34_Park21,17_2024emovomedataset,9_cahyawijaya2023}.

These concerns directly informed our study design. By including both Teenagers and Adults aged 55+ and capturing emotional speech in both Danish (native) and English (non-native), we aimed to empirically explore how such biases manifest in the performance of current SER systems.
To guide this study, we posed the following research question: How do age and language affect the accuracy of deliberately expressed emotions in speech, as interpreted by a speech emotion recognition (SER) model?
Our assumptions were: (1) There is a difference between the two age groups (Teenagers vs Adults 55+ in how they express emotion deliberately; and (2) There is a difference between the two groups when expressing emotions in (native) Danish vs (non-native) English.

\section{User study}
To develop an understanding of people’s ability to deliberately express emotions, we conducted a user study, utilising our prototype, which integrated the SER model developed by Wagner et al.~\cite{38_Wagner23}. The study was carried out in two stages (1) a pilot study to refine the methodology, and (2) a final user study aimed at investigating our research question.

\subsection{Pilot study}
\begin{figure}[h!]
\includegraphics[width=0.6\columnwidth]{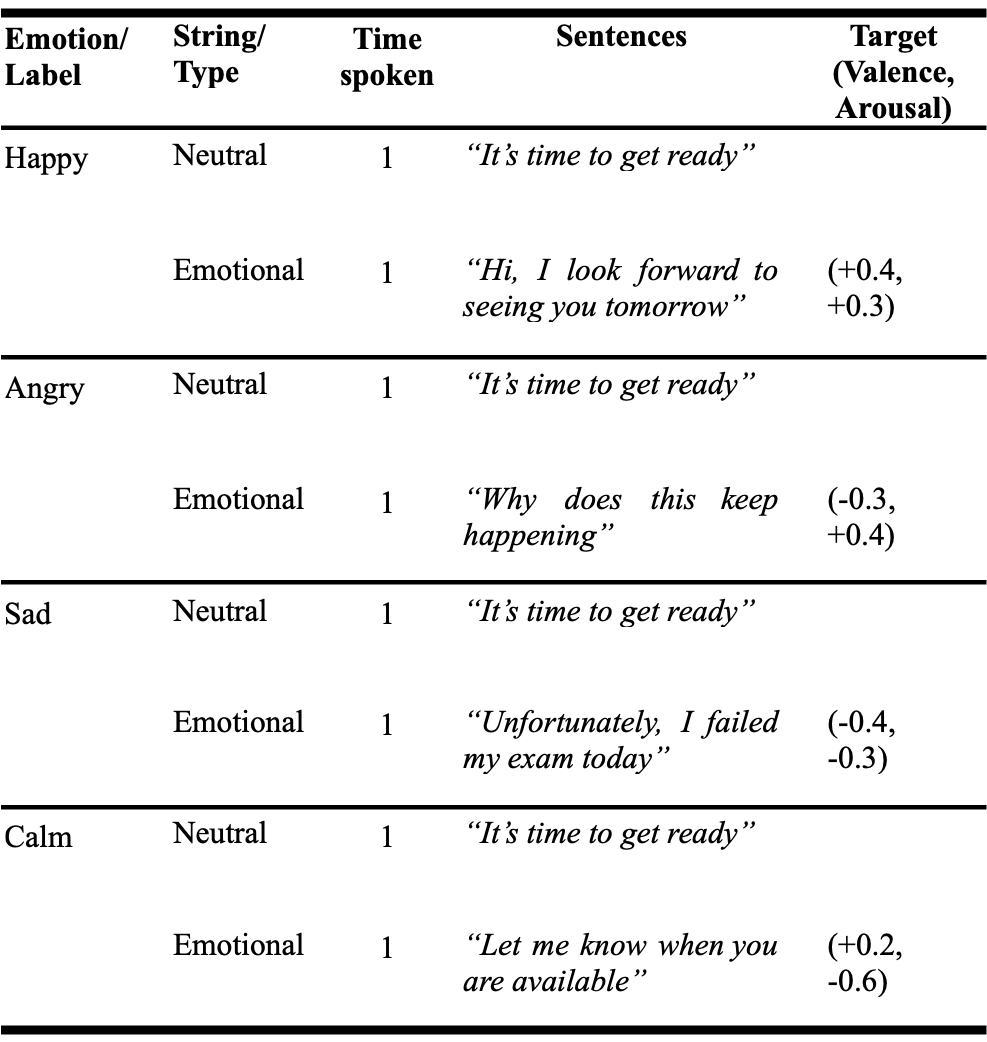}
\caption{The sentences that were uttered in each emotional
quadrant of our model in the pilot study.}
\label{fig:table1}
\end{figure}

To refine the methodology and assess the feasibility of our
user interface (UI) and testing protocol, we conducted a
pilot study with 12 participants from the 55+ age group. All
native Danish. This pilot study aimed to evaluate whether
participants could consistently express target emotions
using a speech-based interface grounded in Russell’s
Circumplex Model of Affect \cite{37_russell1980circumplex}. 
Although the Circumplex 
Model of Affect allows for a wide range of emotional states
across the valence-arousal (V A) space, we selected a subset
of four ``discrete” emotions (happy, sad, angry, and calm) to
represent each dimension. The decision to limit emotional
categories was made to improve the precision of data
collection, reduce participant burden, and ensure clear
mapping between affective states and model responses,
particularly when working in a second language (English).
Each participant was asked to express a series of sentences
under different emotional conditions. Two types of
sentences were used (i) one neutral sentence for its semantic
neutrality, allowing it to be expressed in any of the four
emotions, and (ii) four emotionally biased sentences, which
were semantically biased to fit into one of the four
spectrums of the model. The sentences that were uttered in
the pilot study can be found in Figure \ref{fig:table1}.

\subsubsection{Results from the pilot study} The findings from the pilot
study provided valuable insights that informed the final
design for a user study setup. First, participants had
difficulty pronouncing certain English words in the
emotionally biased sentences, including “unfortunately” and
“available”, which potentially impacted the consistency and
validity of the model’s prediction. Additionally, the SER
model did not show notable differences in detecting
emotions between semantically neutral labels versus
emotional biased sentences. These insights led us to remove
emotionally biased sentences entirely for the final iteration
of the user study.

\subsection{User study setup}
\begin{figure}[h!]
\includegraphics[width=0.6\columnwidth]{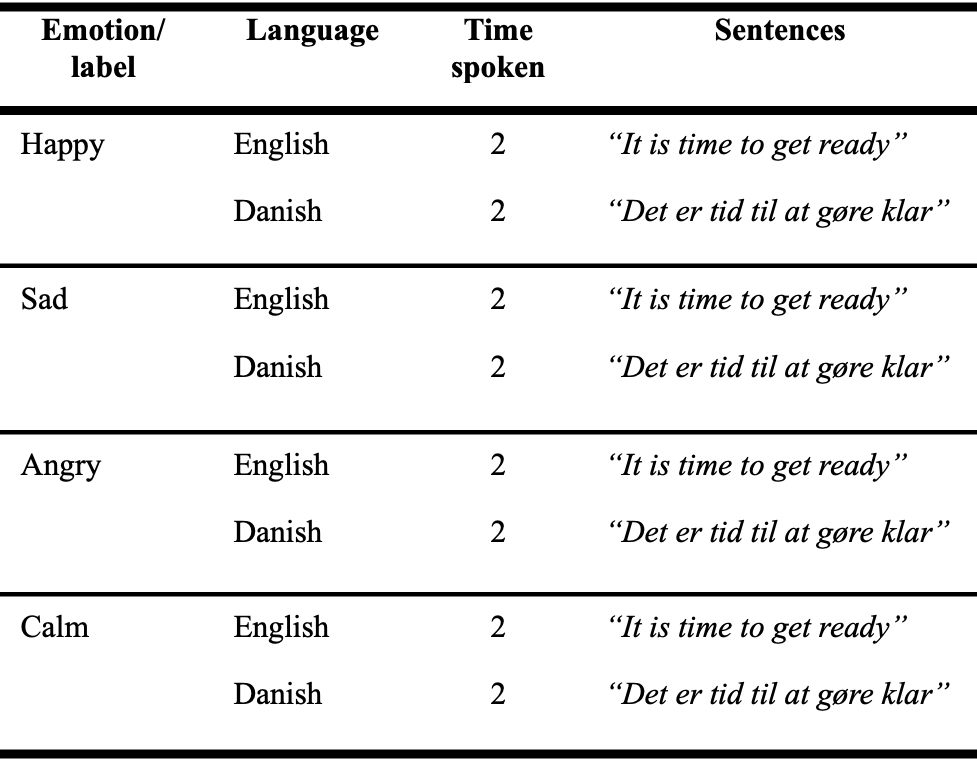}
\caption{The sentences that were uttered in each emotional
quadrant of our model. }
\label{fig:table2}
\end{figure}

The user study setup was designed based on the insights
received from the pilot study. To evaluate potential biases
related to age and language, participants were asked to
deliberately express four target emotions (happy, sad, angry
and calm) using a neutral sentence in both English and
Danish (see Figure \ref{fig:table2}). Each emotion was expressed four times (i) twice in
English, and (ii) twice in Danish. This resulted in 16 total
utterances per participant (four emotions × four repetitions).
To control for practice effects while allowing for consistent
cross-linguistic comparison, the order of emotions was
randomised, while a fixed order was maintained within each
emotional space (English-English-Danish-Danish (EEDD)).

\subsection{Participants}
A total of 24 participants were recruited, with 12 individuals
in each of the two distinct age groups. The teenage group
(five male, seven female), age 15-19 (M = 17), was
recruited from secondary educational institutions in
Denmark, and the Adult group (four male, eight female),
aged 55-82 (M = 67) represented participants with varied
levels of technological familiarity. All participants were
native Danish speakers, an important variable given the
study’s cross-linguistic design, and demonstrated
comparable comfort when speaking English sentences.
Importantly, the inclusion of the 55+ group was intentional,
as older adults remain underrepresented in AI datasets \cite{34_Park21}.
The inclusion of these two groups was designed to reflect a
polarity in both age and technological familiarity \cite{11_chu22, 42_wolfe2024representationbias, 1_abramova2020speech},
which supports the study’s goal of investigating age and
language-related differences in deliberate emotion
expression accuracy.

\begin{figure}[h!]
\includegraphics[width=0.6\columnwidth]{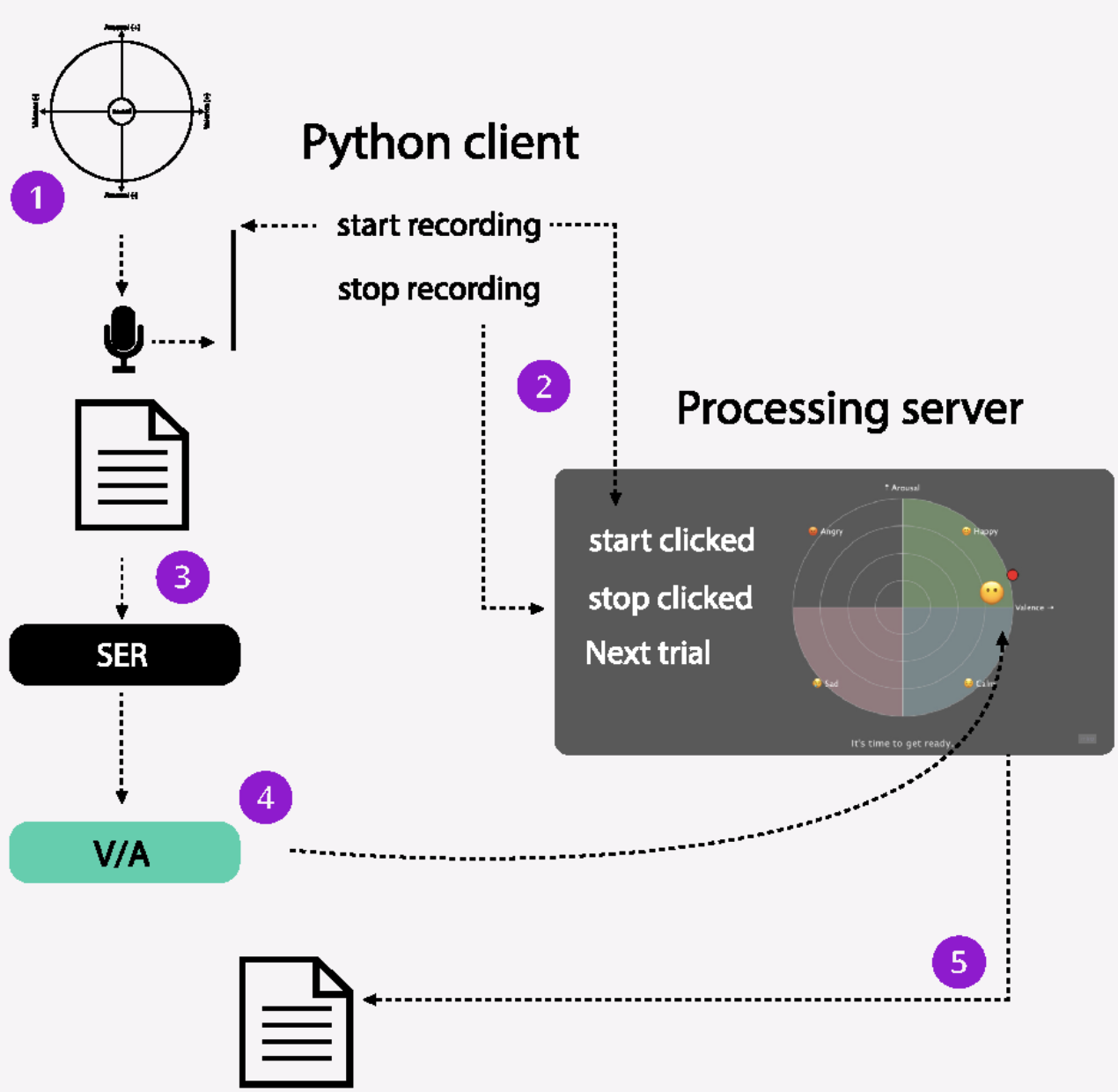}
\caption{(1) Participants indicate their emotional state. (2)
Participants start recording the model. (3) SER results are
mapped to a coordinate. (4). The V/A results are visualised
in the UI (5) Participants are ending the test with an
interview. }
\label{fig:system}
\end{figure}

\begin{figure}[h!]
\includegraphics[width=0.6\columnwidth]{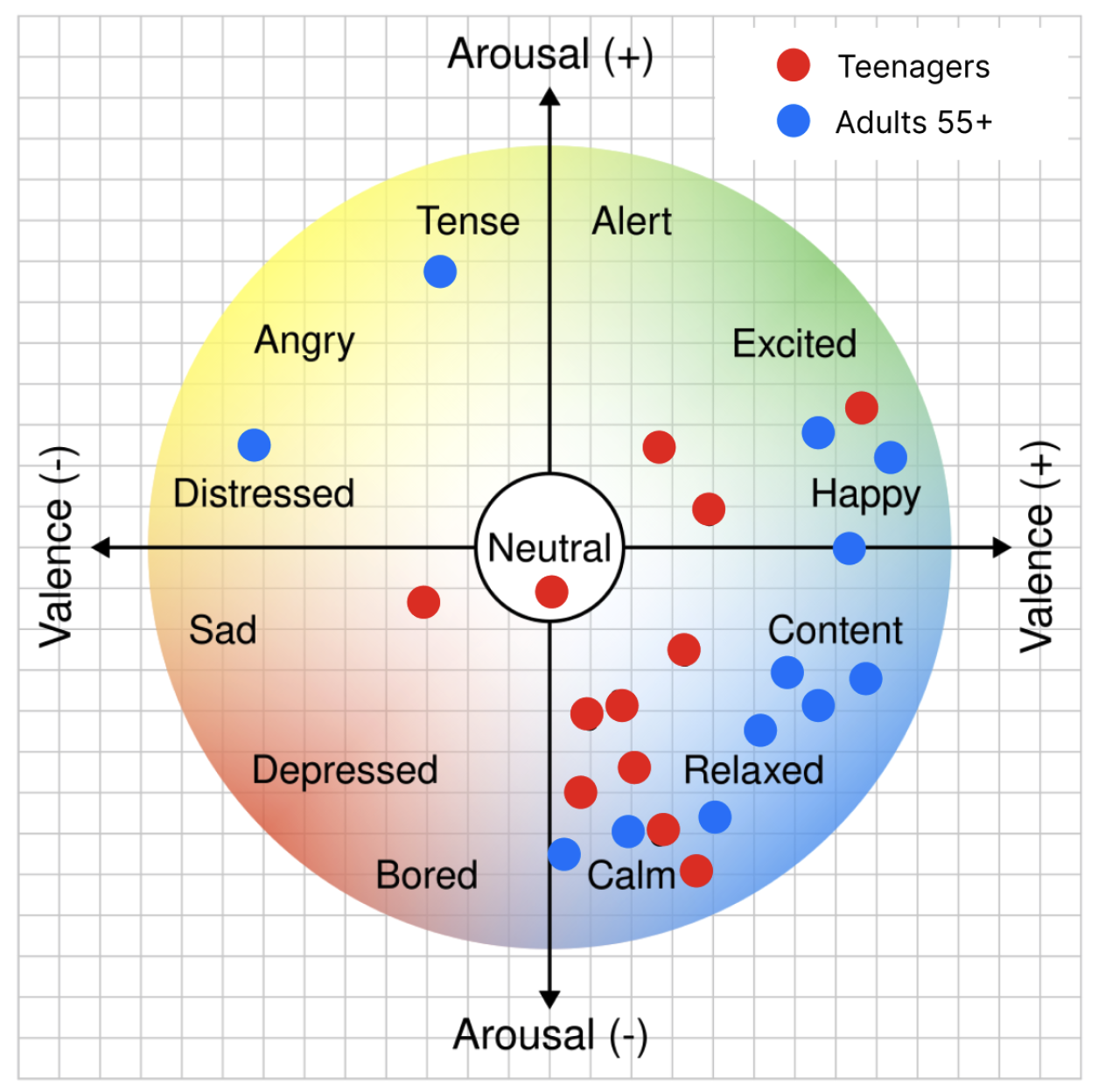}
\caption{Participants' current emotional state before our user
study. }
\label{fig:participants}
\end{figure}

\begin{figure}[h!]
\includegraphics[width=0.6\columnwidth]{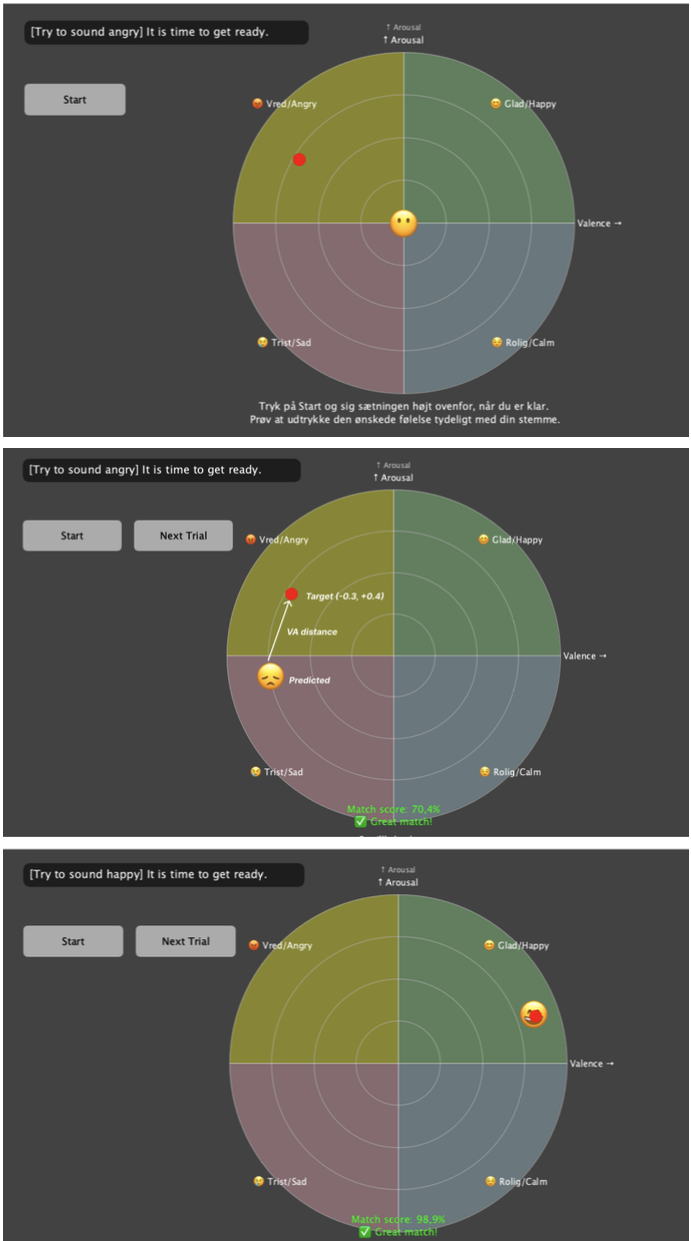}
\caption{Screenshots displaying the user interface used in our
SER-based emotion targeting game, for participants to
interact with. (a) is the start screen, which the participants
see when starting the test (b) showcases an example of the
V A distance measured in the results (c) showcases an
example of a perfect match score, where the expressed
happy emotion closely aligns with the target happy
coordinates. }
\label{fig:procedure}
\end{figure}

\subsection{Procedure}
\begin{figure}[h!]
\includegraphics[width=0.6\columnwidth]{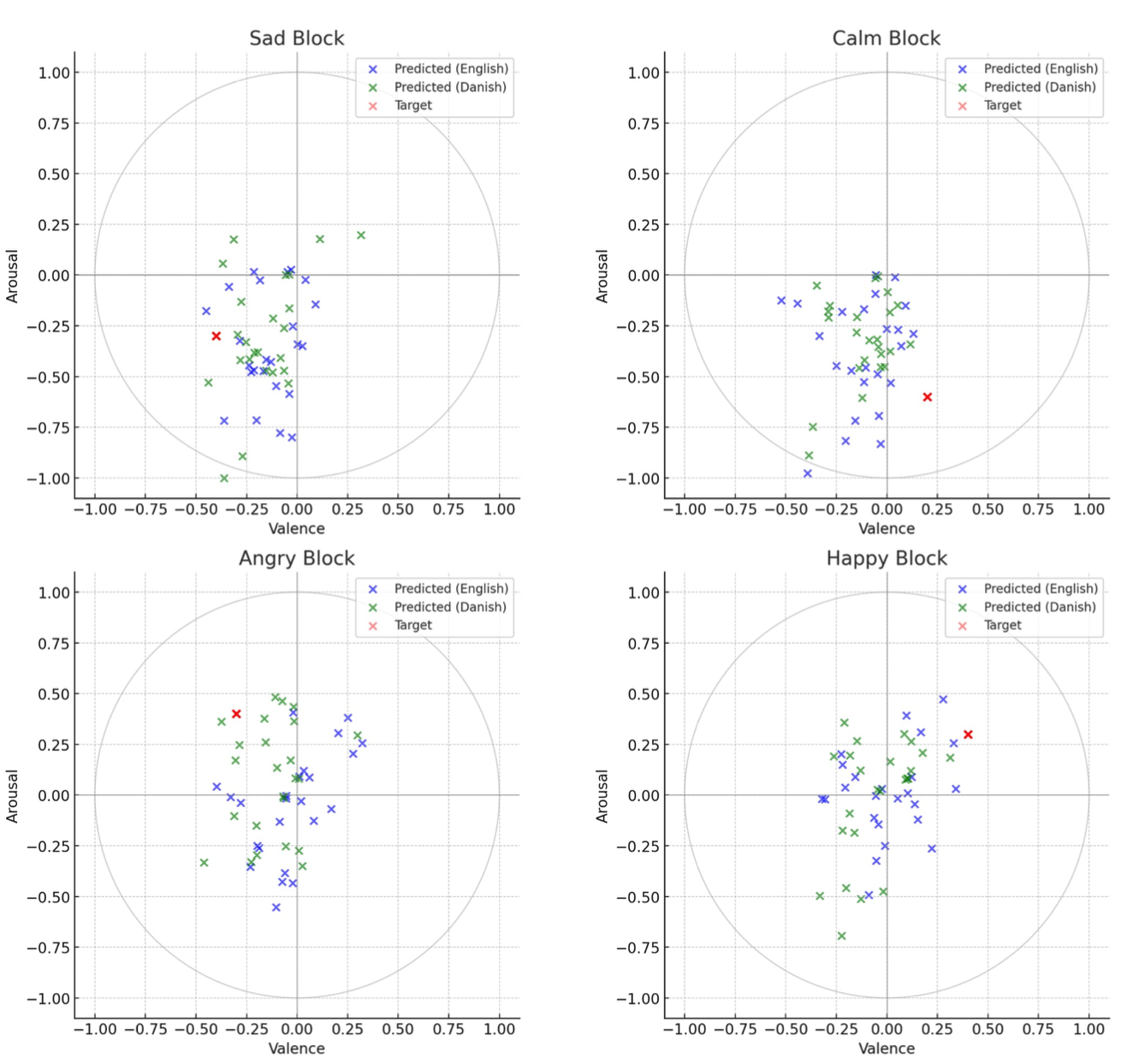}
\caption{Predicted vs. Target Emotion Coordinates: Adults. }
\label{fig:resultsAdults}
\end{figure}

\begin{figure}[h!]
\includegraphics[width=0.6\columnwidth]{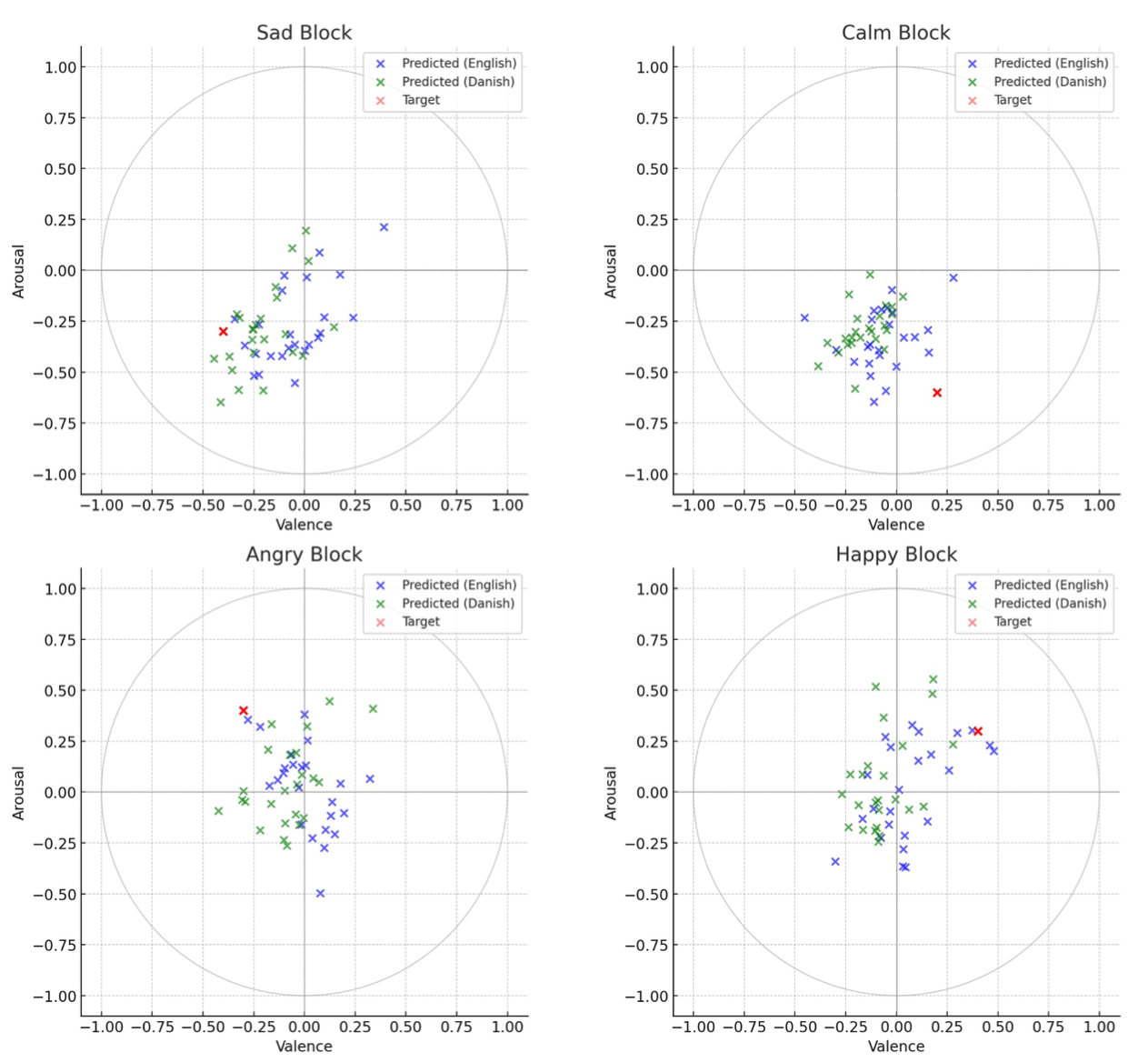}
\caption{Predicted vs. Target Emotion Coordinates: Teenagers. }
\label{fig:resultsTeenagers}
\end{figure}

Each participant completed the study individually. Before
the experiment started, participants were introduced to
Russell’s Circumplex Model of Affect and requested to
identify their current emotional state (see \ref{fig:participants}). Afterwards,
they were introduced to the system interface displaying the
four target emotions (happy, sad, angry and calm), each
mapped to specific coordinates in the valence-arousal (V A)
space, and the sentences which they had to speak (see Figure
\ref{fig:procedure}). They were instructed to vocally express each emotion as
accurately as possible, using the sentence shown on the
screen. This task was conceptualised as a form of emotional
targeting game, where the goal was to ``hit” a specific
emotion coordinate in V A space through their vocal
expression alone. Participants were encouraged to express
each emotion in a way that felt natural and comfortable to
them.
Before official recording began, participants had up to two
minutes to practice using the microphone and familiarise
themselves with our interface. The study was conducted
using a MacBook Air, and participants spoke into the
built-in microphone during both the practice and recording
phases. Once ready, participants proceeded with the
randomised sequence of emotional expression trials in the
fixed language order (English-English-Danish-Danish
(EEDD)). During the
experiment, participants could either control the interface
themselves or have the study administrator manage the
interface (e.g., click at the start, stop, next trial buttons when
needed). After completing the speech recording phase each
participant took part in a semi-structured interview. These
interviews explored topics including the participants'
comfortability with deliberately expressing discrete
emotions in this structured setup context, their experiences
with expressing emotions in a non-native language, general
awareness of SER technology, and their own reflections on
its potential use cases in everyday life applications.
To provide an overview of how the speech emotion
recognition (SER) model interpreted participants’ vocal
expressions, we visualised the predicted emotion
coordinates in valence–arousal space by plotting all logged
coordinates using Python. These plots offer a preliminary
view of the system’s outputs and the degree to which
predictions clustered around the intended emotion targets
across languages and age groups.
Figures \ref{fig:resultsAdults} and \ref{fig:resultsTeenagers} display the predicted valence–arousal (V A)
coordinates for each participant’s spoken utterance, as
generated by the speech emotion recognition (SER) model
and plotted within Russell’s Circumplex Model of Affect.
Figure \ref{fig:resultsAdults} shows data for the Adult group, and Figure \ref{fig:resultsTeenagers} for
the Teenager group. Each figure is divided into four panels
representing the target emotions; sad, calm, angry, and
happy. Predicted coordinates for English utterances are
shown in blue, Danish in green, while red markers indicate
the fixed emotion targets. These plots allow for visual
comparison of expressive accuracy and clustering across
age groups and language conditions, complementing the
statistical analyses in the Results section.

\section{Results}
This section presents the findings from our
user study, which aimed to investigate how age and
language influence the deliberate expression of four
emotions; angry, calm, happy and sad, through speech in
both Danish (participants' native language) and English
(language of model training). We first present the results of
the quantitative analysis, which examines Teenager vs Adult
participants’ expressive accuracy using valence-arousal
(V A) coordinates derived from model predictions across 24
test sessions. This is followed by a qualitative analysis of
participants' interview responses, offering insights into their
subjective experiences with deliberate emotion expression
and their perceptions of speech AI.

\subsection{Quantitative analysis: Expressive Accuracy
in Valence-Arousal Space using SER}
The quantitative analysis draws on 384 emotion-labelled
utterances (16 per participant × 24 participants), evenly split
between Danish and English utterances. For statistical
comparisons and testing, including independent-samples
t-tests, we partitioned the data by language, resulting in 192
utterances per language condition (96 per age group). This
enabled a controlled examination of age-related differences
within each language.
We quantified expressive accuracy as the VA distance (i.e. Euclidean distance) between the model’s predicted and target coordinates in
valence-arousal (V A) space, calculated using the formula: 
\newline
\newline
$VA$ $Distance =\sqrt{(V_{pred} - V_{target})^2 + (A_{pred} - A_{target})^2}$
\newline
\newline
Where $V_{pred}$ and $A_{pred}$ denote the model's  predicted valence and arousal values , and $V_{target}$ and $A_{target}$ the target coordinates of the intended target emotion, based on Russell’s Circumplex Model of Affect \cite{37_russell1980circumplex}. Lower VA
distances indicate greater expressive accuracy.
To test our assumptions (age group differences) and 
(language-related performance differences), we conducted
independent samples t-tests for each emotion within each
language condition. Results for English utterance are
presented first, followed by findings for Danish utterances.

\subsection{English Condition: Age Group Differences
in Expressive Accuracy}
\begin{figure}[h!]
\includegraphics[width=0.6\columnwidth]{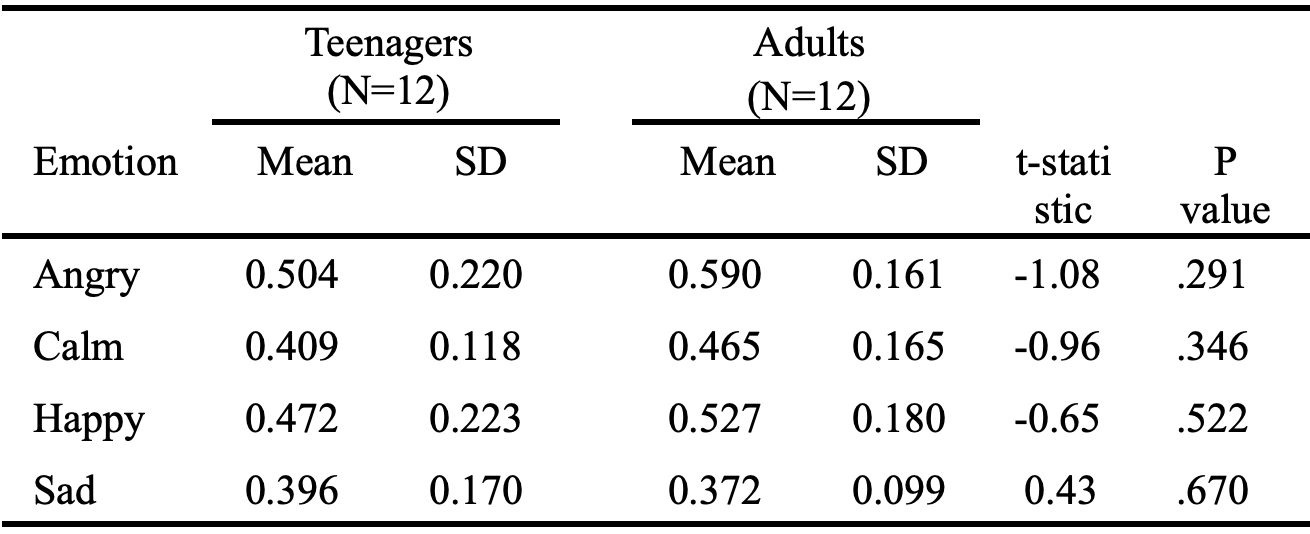}
\caption{Overview of Mean V A Distances by Age Group for
English Utterances.(Note: Lower VA distances indicate greater expressive accuracy. None of the group comparisons reached statistical significance (all p >.05)) }
\label{fig:table3}
\end{figure}

Figure \ref{fig:table3} presents the table with the mean V A distances and standard
deviations for each emotion in the English condition,
comparing Teenagers and Adults 55+.

\begin{figure}[h!]
\includegraphics[width=0.6\columnwidth]{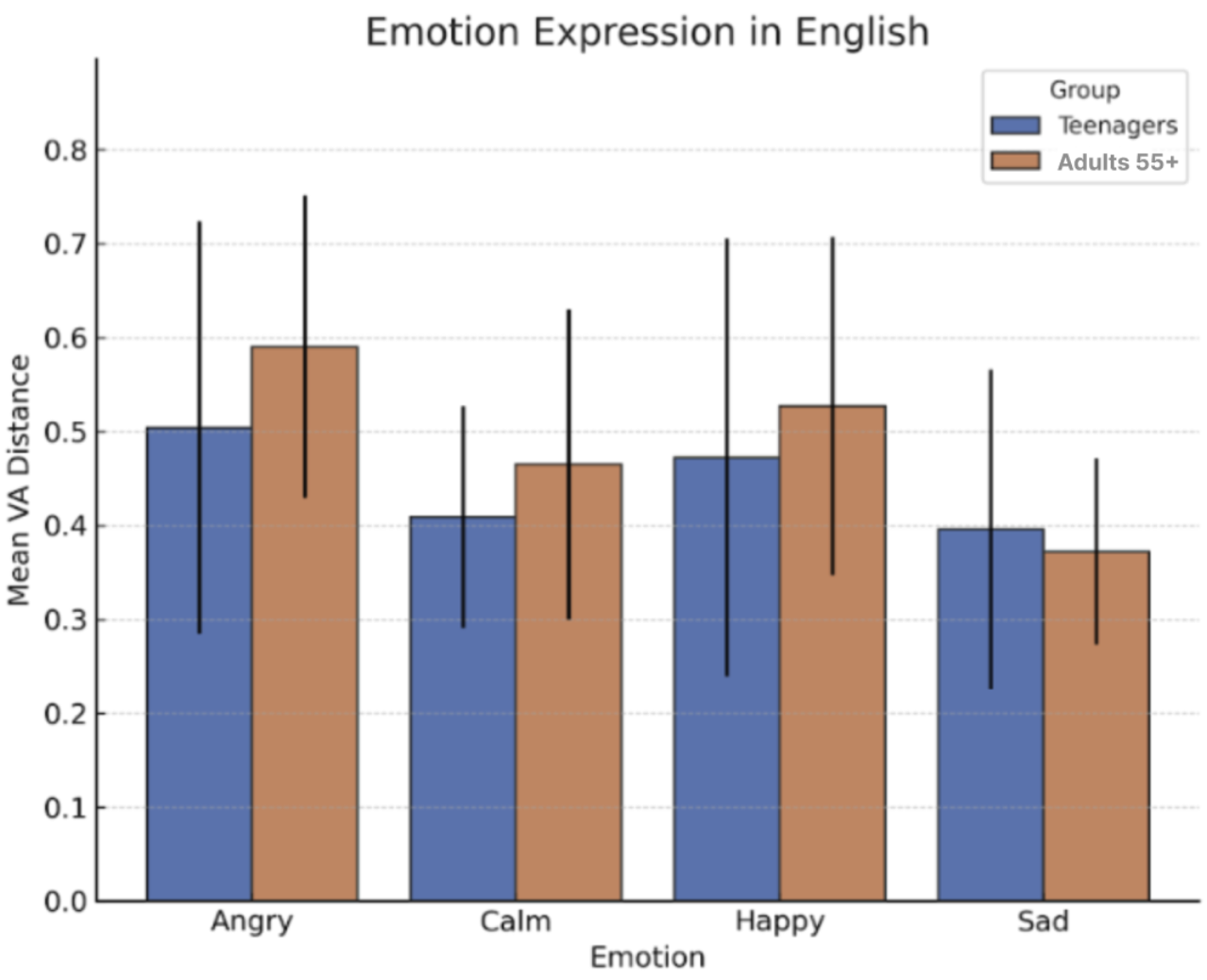}
\caption{Mean V A Distances by Emotion, Age Group and
Language(English). This figure displays mean V A distances
for the four target emotions, comparing Teenagers and
Adults in English (left panel) and Danish (right panel).
Lower V A distances indicate greater expressive accuracy.
Error bars represent standard deviation.}
\label{fig:resultsEnglish}
\end{figure}

As depicted in Figure \ref{fig:resultsEnglish} Teenagers showed marginally lower
mean V A distances than Adults across three (angry, calm,
happy) of the four emotion categories, indicating slightly
higher expressive accuracy in these cases. For angry and
calm, Teenagers demonstrated more accurate vocal
expression (Angry: M = 0.504, SD = 0.220; Calm: M =
0.409, SD = 0.118) compared to Adults (Angry: M = 0.590,
SD = 0.161; Calm: M = 0.465, SD = 0.165).
For happy, Teenagers again exhibited a lower mean distance
(M = 0.472, SD = 0.233) than Adults (M = 0.527, SD =
0.180), reinforcing a pattern in which high-arousal
emotions may be more challenging to express accurately,
particularly for older speakers.
In contrast, sad was the only emotion more accurately
expressed by Adults (M = 0.372, SD = 0.099) than by
Teenagers (M = 0.396, SD = 0.170). However, none of the
differences reached statistical significance (all p > .05),
suggesting broadly comparable levels of expressive
accuracy between age groups in the English condition,
despite small directional trends favouring Teenagers for most
emotions.

\subsection{English Condition: Age Group Differences
in Expressive Accuracy}
\begin{figure}[h!]
\includegraphics[width=0.6\columnwidth]{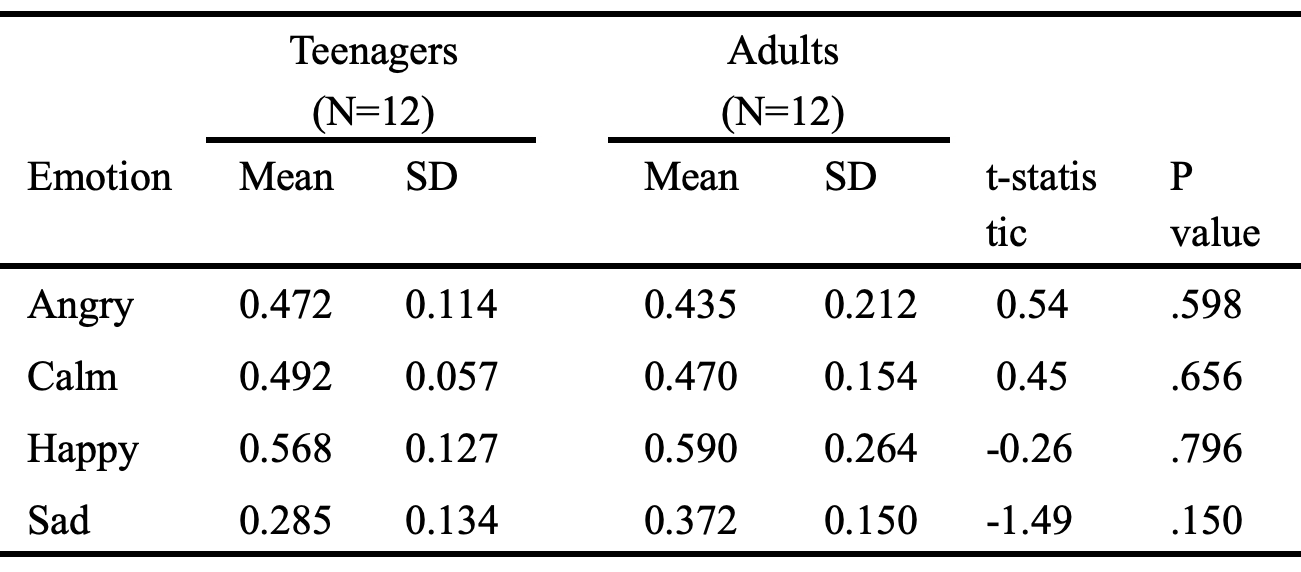}
\caption{Mean V A distances by Age Group for Danish
utterances. (Note: Lower VA distances indicate greater expressive accuracy. The largest group difference (for sad), approached significance (p =.150), but none reached the threshold for statistical significance (all p >.05)) }
\label{fig:table4}
\end{figure}

Figure \ref{fig:table4} presents the table with the V A distance results for Danish utterances, with the same group comparison.

\begin{figure}[h!]
\includegraphics[width=0.55\columnwidth]{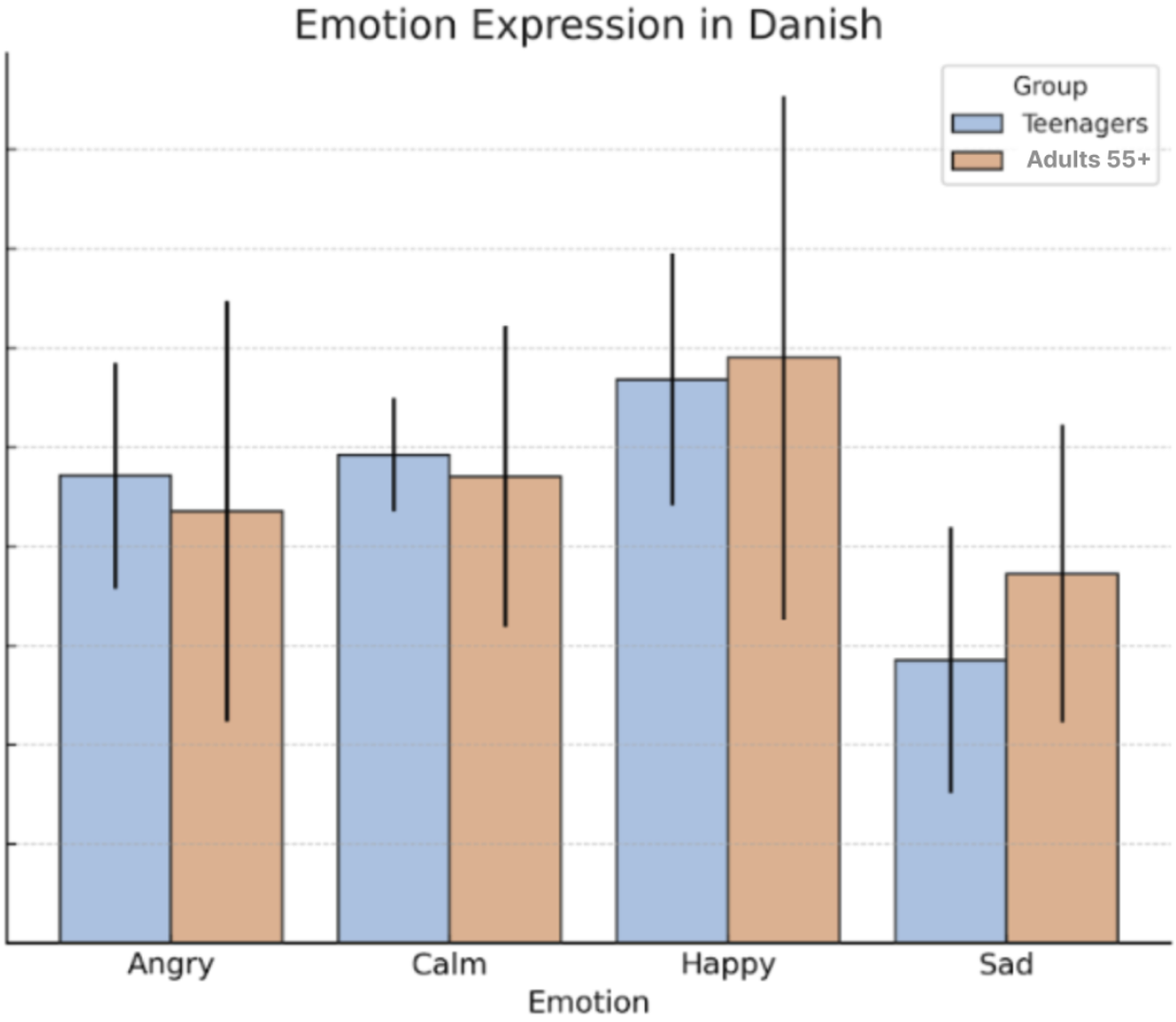}
\caption{Mean V A Distances by Emotion, Age Group and
Language(Danish). This figure displays mean V A distances
for the four target emotions, comparing Teenagers and
Adults in English (left panel) and Danish (right panel).
Lower V A distances indicate greater expressive accuracy.
Error bars represent standard deviation.}
\label{fig:resultsDanish}
\end{figure}

In the Danish condition, group differences in expressive
accuracy varied by emotion, as depicted in Figure \ref{fig:resultsDanish}.
Teenagers demonstrated notably lower V A distances for sad
(M = 0.285, SD = 0.134) compared to Adults (M = 0.372,
SD = 0.150), marking the largest group contrast in the
dataset. Although this difference is not statistically significant
(t(21.74) = -1.49, p = .150). A similar, though less pronounced pattern was observed for happy, where Teenagers again exhibited slightly lower mean
distance (M = 0.568, SD = 0.127) than Adults (M = 0.590,
SD = 0.264). However, for angry and calm, Adults showed
marginally lower mean distances, indicating slightly better
expressive accuracy for these emotions. Adults expressed
angry with a lower mean distance (M = 0.435, SD = 0.212)
than Teenagers (M = 0.472, SD = 0.114), and similarly for
calm (Adults: M = 0.470, SD = 0.152); Teenagrs: (M =
0.492, SD = 0.057).

As in the English condition, happy emerged as the most
difficult emotion to express for both groups, especially
among Adults, who showed the highest V A distance (M =
0.590, SD = 0.264) across all emotion-group combinations.
Overall, while none of the group differences in the Danish
condition reached statistical significance (all p > .05).

\subsection{Qualitative Analysis: Participants
Reflections on Emotion Expression and Speech
AI}
While the quantitative results reveal trends in expressive
accuracy across age and language, they do not capture
participants' subjective experiences. To address this, the
final subsection presents insights from our semi-structured
interviews, highlighting emergent participants' reflections
on the challenges of speech emotion expression and their
attitudes toward AI speech emotion recognition systems.

\subsubsection{Deliberate emotion expression as a task}
Participants across both age groups described the task of
deliberately expressing emotions through speech as
generally challenging, often characterising it as “unnatural”
or “forced”. Teenagers, in particular, tended to use more
emotionally charged or discomfort-oriented language,
describing the experience as “strange”, “awkward”, or “like
crossing a boundary”. Their reflections conveyed a sense of
tension around the performative nature of the task. In
contrast, participants aged 55 and older appeared more at
ease reflecting on their engagement with our system. While
they also acknowledged the artificiality of the task, they
were more likely to frame it as theatrical or performative
rather than purely uncomfortable. One participant shared:
“I’m not an actor, but you try your best - it was ok for me to
deliberately express an emotion.” (P12). Another
participant reflected, “When I attempted to navigate [the
task], it was predominantly more theatrical. It was evident
that it held significant importance, particularly if it was
excessively played.” (P06). Despite shared challenges
across age groups, individuals in the 55+ category were
more likely to describe the task experience as “okay”,
“educational”, or even “fun”.

\subsubsection{Deliberate emotion expression in everyday
contexts}
During the interviews, participants were asked to
reflect on when they deliberately express emotions through
speech in daily life. Almost all participants in the teen group
told, that they were altering the pitch in their voice around
friends all the time. Two Teenagers stated that they changed
pitch when a friend was upset and they wanted to support
the friend, whereas five Teenagers highlighted the value of
changing pitch when informing a friend that they were
feeling sad or angry because of something the friend had
done to them. “Instead of pretending you're fine, you use
your voice to sound more unhappy or angry…” (PL).
Unlike Teenagers, Adults were more likely to shift pitch to
show empathy for someone who was sad and needed
comfort. P11 remarked: "If there's another person who is
feeling sad, then you can try to make them feel happy, by
sounding more cheerful and positive yourself, in a way so
you don't come across as negative to others, for example,
sounding angry unnecessarily". Furthermore, individuals
aged 55+ mentioned that they altered their pitch to convey
professionalism during presentations to colleagues or when
demonstrating authority over a pet or children “If something
unexpected happens and the cat does something, I shout out
loudly” (P7). Another participant commented that they
changed their voice to deliberately convey emotions in both
professional and private settings. “Yes, I do that in some
work situations to put some pressure on those I talk to. Also
with the grandkids” (P8).

\subsubsection{Perceptions of age differences in expressive
accuracy}
Participants were asked which age group they
believed would be more accurate in deliberately expressing
emotions in a system like the one used in our experiment.
Several linked performance to familiarity with and openness
toward technology, suggesting that younger participants
might be at an advantage. One participant, an Adult,
humorously remarked, “There is probably an incorrect
number on my birth certificate” (X4), implying that
younger users may outperform older ones due to
generational differences in tech engagement. Another
reason for participants predicting that teenagers would gain
higher test results was teenagers claiming that Adults’
speech were more monotonous and drier “...especially with
my grandparents, it can be very monotonous, the way they
talk” (PD). A few teenagers further believed that they would
perform better than Adults because of life experience
“Adults have already experienced what we are going
through, and no longer have the same feelings as we do”
(PB). However, participants in both groups stated that they
did not believe that there was any difference in performance
related to age but argued that they thought that the
performance was more related to a specific emotion “I
believe that the older you become, the more relaxed you
get…” (PI).

\subsubsection{The easiest and hardest emotions to convey}
Participants reflected on which of the four target emotions
they found easiest or most difficult to express during the
test. Both age groups generally responded that it was
difficult to deliberately express an emotion they were not
feeling at the time of the test. One Teenager noted that “I
have had a pleasant day, making it difficult to identify a
reason to express sadness” (PD), adding that sounding
angry felt even more difficult under the test circumstances.
Several participants highlighted the feeling of forcing an
emotion as “I thought that was difficult. Of course, you can
convey your emotions, at least in your own head. But we're
all not trained actors.” (P03) Supported by another saying
hat “If I have to sound happy then I also need to be happy”
(P2). The majority of participants in both groups perceived
it as easiest to convey happy or calm, followed up as the
feeling they most frequently expressed towards others. One
of the participants expressed that “It was probably the
emotion I am feeling now. Happy and calm” (X3).
Explaining his personality, one participant described the
emotions he expressed the easiest as “I am a positive and
usually happy person, so it was the emotions in the positive
end that were the easiest.” (X4).

\subsubsection{Understood by the system}
When asked if the
system could accurately convey the participants' emotions,
there was no explicit answer. Participants in both groups
were equally indicating whether their emotions were
correctly represented or not by the system. However, several
participants commented that despite being understood, they
experienced some issues regarding the V A space, pointing
out that emotions close to each other in the Circumplex
Model of Affect (happy/angry or sad/calm) were conveyed
as the opposite emotions due to similarity in the pitch. “It
was relatively good at understanding me when I had to be
happy and angry. It was also good when I was in low
arousal, but in valence, when I had to be calm and sad,
there was a bigger difference than I expected." (PD). This
participant commented that the system found it difficult to
reach the far ends of the valence space. Another participant
did not feel like the system was able to decode their
emotions “Well, I thought I was sounding happy and tried to
speak in a happy way, but it thought I was sad. And when I
tried to sound angry and raise my voice, it said I was sad or
calm. It's not so good with this parameter" (PF). A third
participant criticised the system by commenting how there
are different ways emotions are expressed and how the
system may not be able to decode all aspects of emotions.
“It's very different the way I try to express emotions
compared to how other people express them. So it's evident
that the system cannot decode everyone. People sound
different when they're sad, so it can be difficult for the
system to figure out the whole range of every emotion"
(PH).

\subsubsection{Human VS machine communication}
Participants
explained that the difference between talking to a human
and a machine was that a human possesses body language.
For example, one teen said [..] “I believe that emotions are
closely related to body language, and a system like this
cannot see my body language! It's easy to express emotions
through the body”. Followed up by examples of how it felt
more natural speaking to a human vs a system as a less
forced situation “When you are with another person, you
are unconsciously aware of what you are doing. Then it
becomes a natural reaction, rather than something you do
because you're acquainted to” (X3).

\subsubsection{Familiarity and Trust in SER Technology}
Participants expressed a general skepticism toward fully
automated emotion recognition systems. While
acknowledging the potential utility of SER, many
emphasised the need for a human intermediary, particularly
in high-stakes or professional settings. Seven participants
explicitly suggested that SER technology would be better
suited as a support tool for human specialists rather than a
standalone decision-maker. Interestingly, the nature of
concerns differed by age group. Teenagers were more
focused on the societal implications of widespread AI use,
including fears of over-integration. The Adults were more
critical of the practical implementation and accuracy of the
technology. One participant commented on how it may be
better at diagnosing specific illnesses but should not be used
for consultations; "It could be used within the health sector
in some cases. Some research points to the fact that AI can
be better at detecting diseases than doctors. But that's
different from consultations. " (P3).

\subsubsection{Need for a human third party in Speech AI}
is a
theme that summarises responses to Speech AI related
questions. While participants acknowledged the potential
benefits of SER technology, their comments frequently
implied a discomfort with a fully automated emotional
application. This concern surfaced implicitly as a
recognition of the complexity and subjectivity of human
emotions, which the participants did not feel could be fully
captured by algorithmic means alone. Instead, participants
proposed that the system would require human verification
to make sure that the emotions are perceived correctly by
speech AI technology. One participant explicitly mentioned
the need for a ‘third person’; "I think this could be used
within the health sector. You hear about doctors from the
emergency service who turn away patients. I would like for
the conversations to be recorded, as they are now, but with a
'third person' listening" (P8). Another participant shared
their thoughts on how the system should not be without
human interference; "Artificial intelligence should be
supported by human intelligence. You have to have a second
(human) opinion in different diagnosing systems. At least in
the end. You might be able to filter some things, but after
all, people need to be included" (P2).

\subsubsection{Blending emotions}
is a theme that could be
observed throughout responses about emotion expression.
11 participants spontaneously described emotions as
combinations rather than single states, using examples such
as calm/happy, calm/sad, happy/angry and angry/sad. These
responses indicate that emotions were often perceived and
communicated as overlapping pairs, with several emotions
linked to a particular experience or situation. For example,
one of the participants from the 55+ group described the
most difficult emotions to express as the polarised ones “It
was perhaps the extremes, angry and very happy. It's hard
to say, but it's clear that the voice affects other tones… For
example, the recipient might overreact a bit if your tone is a
little sharp, even if you don't mean it that way, maybe
because you're stressed yourself. So the recipient might
overreact a bit because the voice isn't exactly soft and
pleasant. Speaking in a sharp tone, in a blunt voice towards
others, doesn’t lead to good outcomes. People react more
strongly to that than one might think.” (P12)

\subsubsection{Generational reflections on technology}
Both Teenagers and Adults responded effectively to questions in
the section about emotional expression. However, a
distinction emerged in how each group approached
questions about SER technology. Adults were more likely to
provide reflective and detailed answers, often considering
social or ethical aspects of the technology. In contrast,
Teenagers tend to focus on brief, loaded answers regarding
their perception of SER technology. Although this study
does not explore the underlying tendency of this pattern, the
data indicates a generational gap in the depth of reflection
when discussing AI technology. This observation was
consistent across various interviews, but whether this is
connected to generation, society, technology experience or
something completely different cannot be concluded from
this specific interview.

\section{Discussion}
With increasing integration of SER into everyday life
including healthcare, mobile applications and digital
communication \cite{6_aslan2025speejis, 5_aslan2025speechcommand,3_AnPengcheng2024, 21_hansen2022generalizable}, our study sought to investigate
how age and language affect the deliberate expression of
four discrete emotions through speech, using both Danish
and English utterances. The following discussion analyses
our findings based on the participants’ expressive accuracy
quantified as the Euclidean distance between
model-predicted and target coordinates in V A space, using a
pre-trained SER model.
We argued in the beginning of this paper that SER models
using the machine learning architecture Wav2vec 2.0, which
are trained in monolingual English language corpora, might
be constrained in their effectiveness when applied to other
languages, including Danish. We thereby hypothesised that
the model would exhibit greater alignment with English
sentences, given its trained exposure to language specific
phonetic and prosodic cues from English podcasts \cite{38_Wagner23, 41_Wani}.
However, a particularly noteworthy aspect of our findings
was the cross-linguistic robustness of the SER model we
employed. Contrary to our assumptions, no
statistically significant differences were found between the
age groups or across language conditions(all p > .05). These
results suggest that, under structured elicitation, both
Teenagers and Adults are comparably effective at
deliberately conveying emotions through speech, at least as
interpreted by this model. Moreover, these findings indicate
that the model is capable of identifying emotional states, at
least when measured through paralinguistic cues such as
valence and arousal. This observation is supported by a
study from Pell et al. who proposed that emotional
cues may be partly language independent and partly relying
on universal acoustic features \cite{35_pell2009}. Hence the acoustic
consistency observed in our data suggests a tendency of
capturing emotions to some degree, it does not eliminate the
need to critically investigate how the model performs in
understanding language and cultural interpretation of
emotional cues. Moreover, this also challenges assumptions
about language-specific constraints in affective computing
and raises compelling questions about the universality of
affective prosody.
A recurring theme in our semi-structured interviews was
participants' perception of emotions as blended rather than
discrete, which challenges the ability of SER models to
accurately capture emotion expression using fixed emotion
categories \cite{10_catania2023speech}. While we are not the first to acknowledge
the co-occurrence of same-valence or mixed-valence
emotions \cite{20_grossmann2016emotional,22_israelsson2023blended}, the spontaneous descriptions of blended
emotion states by 11 participants in our study indicates a
more nuanced and multidimensional experience of affect.
This observation highlights that emotions are not always
black and white, with humans feeling one discrete emotion
at the time. With spontaneous conversations in everyday life
emotions are blended and need to be more included in SER
models. This complexity stands in contrast to the static
frameworks that underlie many SER models, which
typically rely on fixed emotion labels \cite{10_catania2023speech, 28_Luo2024}. The
participants' reflection of blended emotions in our study
may reflect the psychological texture of everyday life,
especially in spontaneous communication such as sending
voice messages.
A fundamental limitation in both our study and current SER
practice lies in the reduction of emotional expression to
fixed valence–arousal (V A) coordinates. While this
dimensional model enables structured comparison and
computational efficiency, it oversimplifies the fluid and
context-dependent nature of human affect.
In the controlled clinical setting of our study, this tension
between real-world emotion complexity and categorical
modeling became particularly evident in participants'
difficulty conveying high-arousal emotions like happy and
angry. Our model analysis revealed consistently higher V A
distances for these two emotion categories, suggesting not
only acoustic challenges in modulating these emotions, but
also conceptual ambiguity in their expression. These
findings imply that the dimensional V A space may require
expansion or recalibration to better accommodate these
blended or context-dependent emotional states. This
underscores the need for next-generation SER models to
reflect the emotional realities of humans as diverse users
and reduce systemic bias embedded in speech emotion
recognition \cite{12_Cowie, 31_Brent}.
Interestingly, the results of the semi structured interviews
revealed a notable distinction between subjective participant
self-report and model output. When asked which emotion
participants found easiest to express, most responded with
happy. This self-perception may be shaped by social norms
or the common association of happiness in everyday
situations. However, the model’s analysis indicated that
participants were more accurate when expressing the
emotion sad; in Danish for Teenagers (M = 0.285, SD =
0.134) and in English for Adults (M = 0.372, SD = 0.099).
In contrast to participants' own perception, the emotion
happy consistently elicited the highest V A distances,
especially among Adults (Danish, M = 0.590, SD = 0.264).
This divergence between subjective human-perception and
model-based interpretation suggests that emotional
self-perception and computational results from
paralinguistic cues do not always align.
Notably, Teenagers' strong performance in expressing sad,
may have been influenced by their awareness of how this
emotion affects deliberate vocal pitch. As mentioned in the
semi-structured interviews, several Teenagers described
how they deliberately changed their pitch when feeling sad,
due to emotionally significant situations involving close
friends. This awareness of how their emotional state
manifests vocally may have contributed to their
effectiveness when expressing sad during this study.
Moreover, our findings indicate that conveying emotions
successfully in a model-oriented context is not completely
linked to age but may be influenced by factors such as
personality traits or prior vocal training. For example, one
participant with acting experience outperformed others
across all four emotion labels. This observation
aligns with the participants’ own reflections that
deliberately expressing emotions, particularly in a structured
setting requires techniques used in acting. Such insights are
important in training corpora on fixed labels or spontaneous
speech and emphasise with earlier research in labelling
emotions in (SER) models \cite{10_catania2023speech}.
From a human-computer interaction (HCI) perspective,
these results highlight the need for a human-centric
approach to speech emotion AI. Rather than forcing users to
adapt their expressiveness to algorithmic expectations, we
argue SER systems should adapt to human variability in
emotion expression across age, language and context. By
starting with an understanding of how different people can
express emotion, we aim to place the human, rather than the
algorithm, at the center of future SER models.
Ultimately, SER systems should adapt to the nuances of
human expression, not expect humans to adapt to the
system. To ensure emotional communication is preserved,
we must avoid a future where emotions are lost in
translation - not only between languages, but between
humans and the AI systems meant to understand us. This
study contributes a foundation for building more
emotionally intelligent speech AI that is not only technically
accurate but also ethically aligned and human-centric.

\section{Limitations and Outlook}
While working with emerging technologies such as SER and AI systems designed for user-centric interaction, we encountered several limitations that might have influenced this study. One key limitation relates to the collection of
speech data in naturalistic settings, where the presence of background noise or unintentional vocalisation could not be fully controlled. Additionally, the system’s behaviour during silent input presented a technical limitation. In instances where no sound was detected, such as when a participant hesitated or delayed speech, our model defaulted to the "Sad” quadrant of the valence-arousal space. This default mapping risks introducing interpretive bias, as silence or
hesitation may not necessarily correlate with being sad. If we had given the participants more time then we did to familiarise themselves with the system and conducted the user study in a more controlled environment, the results might have been different. Finally, the study focused exclusively on the valence and arousal dimensions of affect, omitting the dominance dimension commonly included in circumplex
emotion models \cite{37_russell1980circumplex, 36_rizhinashvili2024enhanced}. This scope decision limits the ability to evaluate emotional expressions influenced by
social or hierarchical dynamics, which may be relevant in interactions involving power asymmetries such as age differences.
Given the observed performance limitations, future SER systems may benefit from targeted improvements in recognising high-arousal and both negatively and positively
valenced emotions. Improving the recognition of discrete emotions including happy, sad, calm and angry may be valuable to further investigation. Given their cross-cultural stability and recognisability \cite{15_ekman1992there, 16_Ekman, 10_catania2023speech}, these discrete categories may offer a promising foundation for designing more generalisable SER systems, particularly in real-world
applications where consistent detection is critical \cite{38_Wagner23, 21_hansen2022generalizable}. Emotions, as reflected in naturalistic speech, rarely conform to discrete, predefined labels. Instead, they often emerge as
fluid, blending states shaped by context, personality and cultural norms. For future speech AI systems to more faithfully mirror the psychological complexity of human affect, we argue that they should move beyond binary classifications and toward models that can accommodate ambiguity and emotional nuance. Life, like emotion, is rarely black and white; it unfolds in gradients, tensions and
contradictions - as this experimental study has sought to explore. Acknowledging this theoretical constraint invites further exploration of richer emotion representations, including the dominance dimension and context-dependent
models that better reflect the subtleties of human affective
experience. Such lines of future inquiry may also benefit
from attention to spontaneous vs acted expressions that have
shown to significantly affect recognition outcomes. By
targeting both emotional complexity and universal patterns
of expression, future research can better align SER
technologies with the nuances of human communication and
emotion detection.
Another pattern of note relates to how the SER model
prioritised paralinguistic over semantic cues. Despite the
SER model being trained exclusively on English-language
data, participants’ Danish utterances were interpreted with
comparable expressive accuracy. This suggests that certain
affective signals, such as pitch and vocal energy, may be
universally encoded across languages, grounded in shared
physiological and emotional mechanisms. Considering that acoustic-linguistic speech emotion recognition can suffer from word errors in automatic speech recognition (ASR) \cite{amiriparian2021}, being able to mainly rely on the acoustic cues for speech emotion recognition would be a technically desirable situation. During the user
study, several participants began laughing while trying to
express the “angry” emotion, finding it amusing to simulate
the emotion angry on command. In these cases, the SER
model interpreted the laughter acoustically and visually as
“happy,” overriding the intended angry target. These
observations affirm that voice tone can be more salient to
SER than the actual words spoken, highlighting how
paralinguistic cues can dominate the model’s interpretation.
Building on these observations, one interesting forward
looking direction would be to apply our experimental design
to structurally underrepresented languages such as
Greenlandic, which was only recently added to commercial
translation platforms Google Translate \cite{18_google_translate_2024}. Given its
phonetic and prosodic uniqueness, evaluating SER models
across such typologically distant languages could help
illuminate whether emotional prosody is indeed universal. If
an English-trained model can decipher emotional cues in
Greenlandic utterances, it would suggest that the
grammar of emotion is more ancient and cross-cultural than
the grammar of language itself, a hypothesis with profound
implications for the next generation of affective computing
and cross-cultural AI. We acknowledge that the complex and ambitious, but also pioneering nature of the user study comes with a variety of limitations. We hope that our research will contribute to a critical and carefully application and design of interactive speech emotion AI solutions.  
Finally, in this preprint we presented a preliminary and explorative analysis of the quantitative data, focussing on mean V A distances. We aim to explore advanced methods and make the dataset public in future iterations.

\section{Conclusion}

Biases in machine learning solutions and SER systems exist and are being studied. For example, fellow researchers have explored gender bias \cite{Lin_2024, gorrostieta2019gender} in SER. The motivation for the study at hand was driven by similar assumptions, including that Teenagers and Adults 55+ are underrepresented in training datasets, potentially contributing to challenges ensuring individual fairness across diverse age-groups. However, we were also interested to study such biases in a realistic and interactive setting with (new) users, where users are aware of the speech emotion AI's functionality and can deliberately modify their speech emotions when interacting with a trained model (both in English the language the model was trained and their native Danish language). We assumed that age and language could shape the ability of such deliberate vocal expression of emotion and impact (either counterbalance or worsen) any bias based on the training data. In response, this paper introduced an interactive prototype and reported on a speech-emotion-acting study in the field with participants in both age-groups and in two languages (Danish and English). We reported in-debt qualitative and quantitative analyses of their opinions, performance and targetted speech emotion acting patterns. The statistical analysis based on independent t-tests and mean (V A) distance as the dependable variable showed no significant effect of age or language on the performance of deliberately hitting SER targets. In summary, the research at hand outlines  potentials of misalignment between the standard valence-arousal (V A) coordinates employed in current SER systems with how emotions are deliberately and  vocally expressed in both user groups and languages. We believe that some of the identified challenges could be addressed through recalibrating emotion coordinates in future SER application to ensure more accurate and inclusive speech AI in the field.



\bibliographystyle{ACM-Reference-Format}
\bibliography{references}


\end{document}